\documentclass[11pt]{article}
\usepackage{graphics}
\usepackage{tikz}
\usepackage{amsmath}
\usepackage{amssymb}
\usepackage{units}
\usepackage{comment}
\usepackage[top=1.25in, bottom=1.25in,
            left=1.7in, right=0.5in, twoside]{geometry}
\usepackage{subcaption}
\usepackage[textwidth=22mm, size=footnotesize]{todonotes}
\usepackage{hyperref}

\usepackage[square,numbers]{natbib}

\newcommand{\ds}{\displaystyle}
\begin{document}

\title{A bi-directional low-Reynolds-number swimmer with passive elastic arms}

\author{Jessie Levillain\footnote{CMAP, CNRS, École polytechnique, Institut Polytechnique de Paris, 91120 Palaiseau, France} \and François Alouges\footnote{Centre Borelli, ENS Paris-Saclay, CNRS, Université Paris-Saclay, 91190 Gif-sur-Yvette, France} \and Antonio DeSimone\footnote{SISSA, Trieste, Italy} \and Akash Choudhary\footnote{Department of Chemical Engineering, Indian Institute of Technology,  208016 Kanpur, India} \and Sankalp Nambiar\footnote{Department of Chemical Engineering, Indian Institute of Technology Delhi, New Delhi 110016, India} \and Ida Bochert\footnote{Max Planck Institute for Medical Research, Jahnstraße 29, 69120 Heidelberg, Germany and Institute for Molecular Systems Engineering and Advanced Materials, Heidelberg University, Im Neuenheimer Feld 225, 69120 Heidelberg, Germany}}

\date{}

%

%
%
\maketitle
\begin{abstract}
It has been recently shown that it is possible to design simple artificial swimmers at low Reynolds number that possess only one degree of freedom and, nevertheless, can overcome Purcell's celebrated scallop theorem. One of the few examples is given by Montino and DeSimone, Eur. Phys. J. E, vol. 38, 2015, who consider the three-sphere Swimmer of Najafi and Golestanian, replacing one active arm with a passive elastic spring. We further generalize this idea by increasing the number of springs and show that it is possible to invert the swimming direction using the frequency of the single actuated arm.
\end{abstract}
%

\section*{Introduction}
\label{intro}

One of the many features of low-Reynolds-number swimming \cite{childress_1981, Becker,Stone} is the {\emph{Scallop Theorem}} stated by Purcell in his celebrated lecture \cite{purcell_life_1977}, which forbids any net motion produced by {\emph{reciprocal}} shape changes of the swimmer. Several mechanisms have been, since then, proposed to overcome the scallop theorem (see \cite{lauga_hydrodynamics_2009} and the many references therein), most of them using  strategies to create non trivial loops, i.e. closed curves with non zero surface, in the shape space of the swimmer. Simple examples are given by the three link swimmer \cite{purcell_life_1977}, the three sphere swimmer \cite{najafi_simple_2004}, the Push-Me-Pull-You \cite{Avron2005PushmepullyouAE} or Purcell's rotator \cite{dreyfus_purcells_2005}. All of these examples use two degrees of freedom in the shape activated periodically in time with a phase lag in order to produce the loop. 

Furthermore, the efficiency of such systems has also been extensively studied in the literature, and we refer the interested to the examples given in references \cite{alouges_energy-optimal_2019, Avron2004, alouges_optimal_2008, Tam}.

Strikingly, the constraint of having at least two degrees of freedom can be overcome, if one considers in the system, a passive elastic link. Indeed, the mechanism, discovered in \cite{passov_dynamics_2012}, consists of a three-link swimmer with only one activatable joint, the other being a rotational spring. Due to hydrodynamic interactions, a periodic activation of the controlled parameter induces an out of phase periodic motion of the spring, permitting to describe a loop in the shape space. So far, two swimmers have been proposed in that direction: the three-link swimmer \cite{passov_dynamics_2012} and the three-sphere swimmer \cite{montino_three-sphere_2015}. The latter, being an
adaptation of the one proposed in \cite{najafi_simple_2004}, is simpler to analyze because, due to its rectilinear shape, no rotational movement is possible. A careful study, provided in \cite{montino_three-sphere_2015} shows in particular that a displacement of the swimmer is possible depending on the frequency at which the actuated arm is deformed. At very low or very high frequency no net motion is possible after a stroke, while at a natural frequency of the system (which depends on the viscosity of the fluid, the masses and the spring constant) a resonant effect takes place which provides an out of phase oscillation of the spring and, consequently, a net motion. 

The present paper explores the situation where another spring is added to the system. The swimmer is therefore composed of four spheres linked by three arms, one of which (in the center) being activated while the two remaining ones are springs. We particularly focus on the situation where the spring constants are different, leading to 2 natural frequencies of the system. We show that such systems can offer the interesting possibility of controlling the direction in which the swimmer moves with the frequency of the activated arm.

After recalling Montino and DeSimone's results on the three-sphere swimmer \cite{montino_three-sphere_2015}, we introduce a two-spring swimmer with a spring on each side of the actuated arm, and explore the consequences on its swimming motion and direction. We  study its dynamics and displacement, depending on various system parameters. 

\section{The 3-sphere Swimmer with a passive arm}
\label{sec:twolink}

\begin{figure}[h!]
\centering
 \includegraphics[width = 0.7 \textwidth]{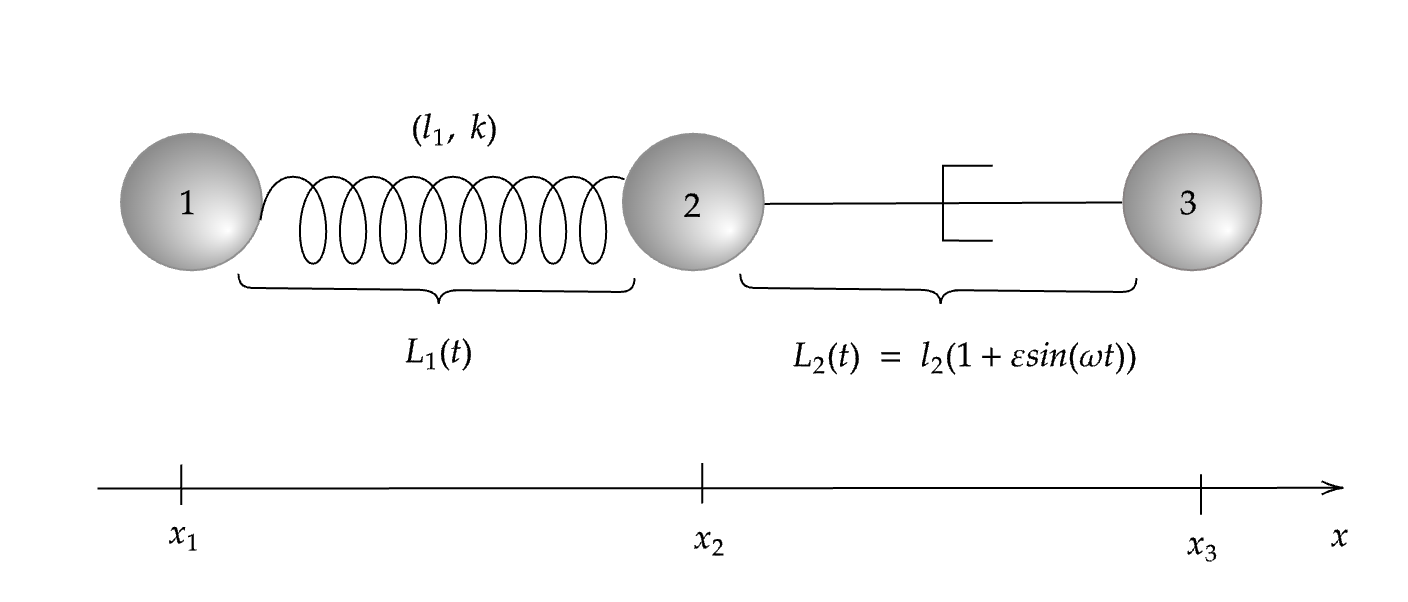}
\caption{Three-sphere swimmer as proposed by Montino and DeSimone in \cite{montino_three-sphere_2015}. Two spheres are connected by a spring with rest length $l_1$ and spring constant $k$. The third sphere is connected by an actuated arm. The distances between the spheres are given by $L_1(t)$ and $L_2(t)$.}
\label{figure:scheme_three_sphere_swimmer_Montino_DeSimone}
\end{figure}

Najafi and Golestanian created a swimmer consisting of three spheres, connected by two actuators \cite{najafi_simple_2004}. By increasing and decreasing the length of the two actuators out of phase, the swimmer is able to perform a non-reciprocal motion providing a net displacement. Based on this, Montino and DeSimone proposed a three-sphere swimmer with a passive elastic arm \cite{montino_three-sphere_2015} mimicking in this context, the mechanism of \cite{passov_dynamics_2012}. As shown in Figure \ref{figure:scheme_three_sphere_swimmer_Montino_DeSimone}, the first and second spheres are connected by a spring with spring constant $k$. The equilibrium length is denoted as $l_1$, the distance between the spheres is $L_1(t)$. The length of the actuated arm, which connects the second and third spheres, is given as $L_2(t)$. The radius of the spheres is $a$. For a periodic activation of $L_2(t)$ of the form 
\begin{align}
	L_2(t)=l_2(1+\varepsilon \sin (\omega t)),
	\label{equation:L2_Montino_deSimone}
\end{align}
the passive elastic arm experiences an out-of-phase oscillation due to hydrodynamic interactions. The movement $L_1(t)$ can be calculated by balancing elastic and hydrodynamic forces. Indeed, assuming $\frac{a}{L_1} \sim \frac{a}{L_2} \ll 1$, the velocities $v_i$, $i=1,2,3$ of the three spheres are related to the hydrodynamic forces in the following way
\begin{align*}
	v_1&=\frac{f_1}{6\pi \mu a}+\frac{f_2}{4\pi \mu L_1}+\frac{f_3}{4 \pi \mu (L_1+L_2)}, \\
	v_2&=\frac{f_1}{4\pi \mu L_1}+\frac{f_2}{6\pi \mu a}+\frac{f_3}{4 \pi \mu L_2}, \\
	v_3&=\frac{f_1}{4\pi \mu (L_1+L_2)}+\frac{f_2}{4\pi \mu L_2}+\frac{f_3}{6 \pi \mu a}.
\end{align*}
Here, $\mu$ is the fluid viscosity and $f_i$ describes the force, which is exerted on the fluid by the $i$-th sphere.  
The above set of equations can also be represented in the matrix form
\begin{equation}
    V = R F
    \label{eq: MF}
\end{equation}
where $V=(v_1, v_2, v_3)^T$ is the vector of the velocities of the spheres and $F=(f_1, f_2, f_3)^T$ the vector of hydrodynamic forces acting on each sphere. The current structure of $3 \times 3 $ mobility matrix $R$ takes into account stokeslet interactions \cite{ref-Batchelor} by assuming that $a \ll L_i$, while self-interaction diagonal terms come from the hydrodynamic motion of a sphere in a fluid.

As it is the case in the low Reynolds number regime, inertial forces are neglected. Since there are no external forces considered, the force balance equation for a freely suspended swimmer
\begin{equation}
    f_1+f_2+f_3=0
    \label{eq: forcebalance_montino}
\end{equation} 
must be fulfilled. The elastic force $f\textsubscript{e}=k(L_1-l_1)$ acting on the first sphere is equal and opposite to $f_1$ due to Newton's third law, leading to
\begin{equation}
    f_1 = k(L_1-l_1).
    \label{eq: elastic_fluid_montino}
\end{equation}

\begin{figure}[ht]
	\begin{subfigure}{0.45\textwidth}
		\centering
		\includegraphics[width=0.9\linewidth]{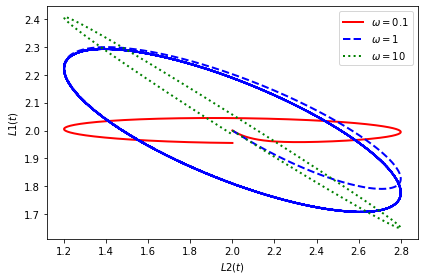}
		\subcaption{}
		\label{fig:montinodesimone1}
	\end{subfigure}
	\begin{subfigure}{0.45\textwidth}
		\centering
    \includegraphics[width=0.9\linewidth]{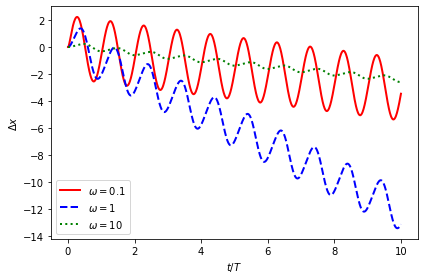}
		\subcaption{}
		\label{fig:montinodesimone2}
	\end{subfigure}
	\caption{Numerical results for Montino and DeSimone's three-sphere swimmer. (A) Curves in configuration space resulting from numerical solution of equations (\ref{equation:L2_Montino_deSimone} - \ref{equation:L2dot_v2_v3_Montino_DeSimone}). The different colors correspond to different frequencies $\omega$ (in $\unit[]{rad ~s^{-1}}$) of the actuated arm. Note that the unit of length is $\unit[10^{-4}]{m}$. (B) Displacement $\Delta x$ (in $\unit[10^{-4}]{m}$) of the three-sphere swimmer with respect to the normalized time $t/T=t\cdot \omega/(2\pi)$ for different actuating frequencies $\omega$. \cite{montino_three-sphere_2015}}
\end{figure}

Furthermore, the velocities of the spheres and the lengths of the two arms are connected by
\begin{align}
	\dot{L}_1=v_2-v_1\,, \quad \dot{L}_2=v_3-v_2\,.
	\label{equation:L2dot_v2_v3_Montino_DeSimone}
\end{align}
The time evolution of $L_2$ is controlled by the actuator as given in equation \eqref{equation:L2_Montino_deSimone}. The given system of equations (\ref{equation:L2_Montino_deSimone}- \ref{equation:L2dot_v2_v3_Montino_DeSimone}) can be solved to obtain $L_1(t)$ and $\dot{L}_1(t)$ and this calculation is done numerically in \cite{montino_three-sphere_2015}. It can be observed that loops are formed in the configuration space  $(L_1,L_2)$. Since the displacement of a swimmer is proportional to the (algebraic) area enclosed by the curve in shape space, the three-sphere body is able to swim. Figure \ref{fig:montinodesimone1} shows the calculated curves in the shape space for different values of the actuating frequency $\omega$. The other parameters are set to $l_1=l_2=\unit[2\cdot 10^{-4}]{m}$, $a=\unit[0.1\cdot 10^{-4}]{m}$, $\mu=\unit[8.9\cdot 10^{-4}]{Pa~ s}$, and $k=\unit[10^{-7}]{N/m}$. It can be observed that the area enclosed by the blue curve (corresponding to an intermediate frequency) is larger than for the higher frequency (green curve) and for the lower frequency (red curve). This agrees with the qualitative study of \cite{montino_three-sphere_2015}, which states that the area enclosed by the curve vanishes for $\Omega \rightarrow 0$ or $\Omega \rightarrow\infty$, where $\Omega = \omega \mu (l_1+l_2)/k$ is a dimensionless frequency parameter. 

From the resulting time evolution of $L_1(t)$ and $L_2(t)$, the position $x_i$ of each sphere can be calculated from $\dot{x}_i=v_i$. The displacement of the swimmer $\Delta x(t)=\sum_i \left(x_i(t) -x_i(0)\right)/3$ is shown in Figure \ref{fig:montinodesimone2}, with respect to the normalized time $t\cdot \omega/(2\pi)$. As it can be observed from the area in configuration space, the displacement of the swimmer after one period with intermediate frequency (blue) is larger than the one corresponding to a smaller (red) or higher frequency (green). 

It can be concluded that the three-sphere swimmer proposed by Montino and DeSimone is able to produce a net displacement depending on the dimensionless frequency $\Omega$ of the actuated arm. If $\Omega$ becomes too high or too low, the net displacement of the swimmer becomes smaller or eventually vanishes.

\section{The 4-sphere swimmer with two passive elastic arms}
\label{sec:2.1}

Montino and DeSimone's swimmer \cite{montino_three-sphere_2015} enabled us to circumvent the \textit{Scallop Theorem} while only having one active degree of freedom in the system. However, this three-sphere swimmer can only swim in one direction. We now propose a new mechanism which, still with a single degree of freedom, is able to swim in both directions.

\begin{figure}[h]
\centering
\includegraphics[width = 0.9 \textwidth]{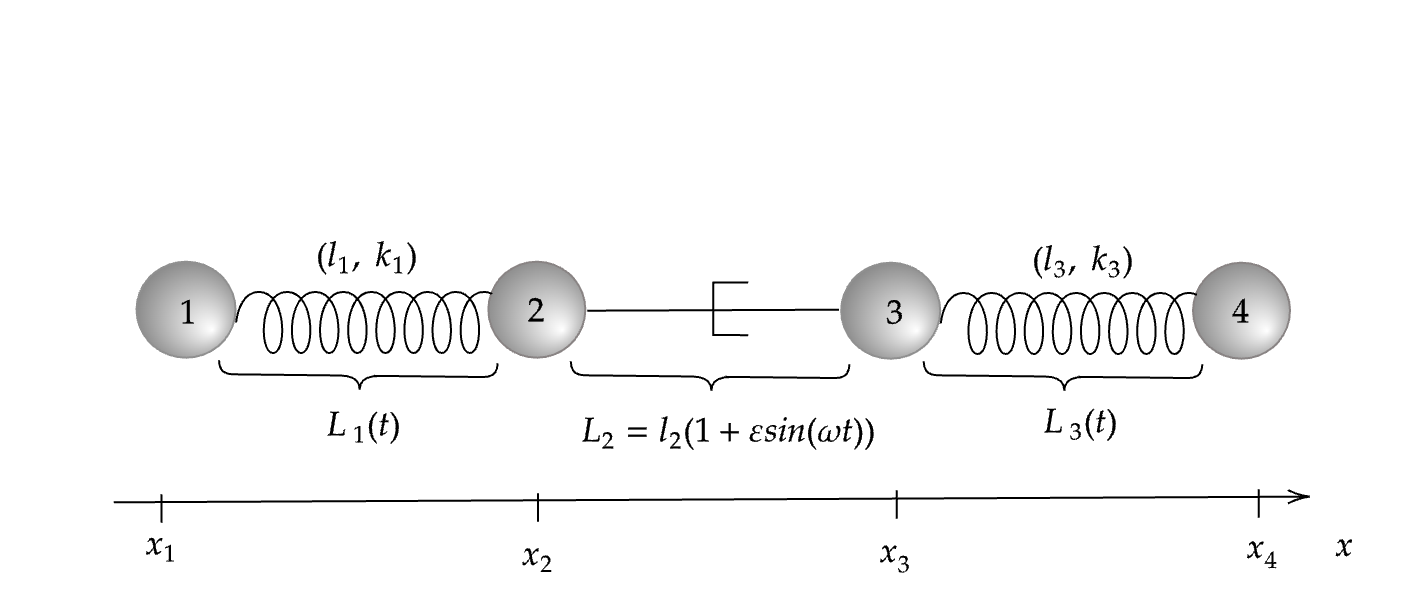}
\caption{Four-sphere swimmer where the middle arm connecting spheres 2 and 3 is the activated part. Springs have rest lengths $l_i$ and spring constants $k_i$. Distances between the spheres are given by $L_1(t)$, $L_2(t)$ and $L_3(t)$. In this case, the motion of the middle arm is prescribed as: $L_2 = l_2(1 + \varepsilon \sin(\omega t))$.}
\label{fig:middle_link_schematic}
\end{figure}

The swimmer is composed  of 4 spheres where only the middle link serves as the active arm while the two others are passive springs as shown in Figure \ref{fig:middle_link_schematic}.
 In other words, 
 \begin{equation}
     L_2 = l_2(1+\varepsilon\sin(\omega t)), 
     \label{eq: activeL2}
 \end{equation}
and $L_1$, $L_3$ are the unknowns to be determined.

 As in Section \ref{sec:twolink}, we write down the equations for velocity $v_i$'s of the four spheres similarly to \cite{montino_three-sphere_2015}. 
This yields the following set of equations that govern the velocity of the spheres:
\begin{subequations}
	\begin{eqnarray*}
		v_1 &=& \frac{f_1}{6\pi \mu a}+\frac{f_2}{4\pi\mu L_1}+\frac{f_3}{4 \pi\mu (L_1+L_2)} + \frac{f_4}{4 \pi\mu (L_1+L_2+L_3)}, \label{eq:three_link_v_a}\\
		v_2 &=& \frac{f_1}{4\pi \mu L_1}+\frac{f_2}{6 \pi\mu a}+\frac{f_3}{4 \pi\mu L_2} + \frac{f_4}{4 \pi \mu (L_2+L_3)}, \label{eq:three_link_v_b}\\
		v_3 &=& \frac{f_1}{4\pi \mu(L_1+L_2)}+\frac{f_2}{4 \pi\mu L_2}+\frac{f_3}{6 \pi\mu a} + \frac{f_4}{4 \pi \mu L_3},\;\;  \label{eq:three_link_v_c}\\		
		v_4 &=& \frac{f_1}{4\pi\mu (L_1+L_2+L_3)}+\frac{f_2}{4 \pi \mu(L_2+L_3)}+\frac{f_3}{4 \pi \mu L_3}+\frac{f_4}{6 \pi\mu a}\,,		\label{eq:three_link_v_d}
	\end{eqnarray*}
\end{subequations}
 that we again write in the abstract form
 \begin{equation}
     V=RF\,,
     \label{eq:velocity-force}
 \end{equation}
where $V=(v_1, v_2, v_3, v_4)^T$ is the vector of the velocities of the spheres and $F=(f_1, f_2, f_3, f_4)^T$ the vector of hydrodynamic forces acting on each sphere.
 Furthermore, the force-free condition becomes 
 \begin{equation}
     f_1+f_2+f_3+f_4 = 0.
     \label{eq:force4}
 \end{equation}
 and we can write the force balance law on spheres 1 and 4 as the balance between the viscous forces and the elastic forces of the arms:
\begin{align}
	f_1 = k_1 \left(L_1-l_1\right),\, \quad \, f_4 = - k_3 \left(L_3-l_3\right)\, .
\label{eq:three_link_force_bal}    
\end{align}
and we observe that the rate of change of each of the links obeys 
\begin{equation}
    \dot{L}_i = v_{i+1} - v_i, \text{ for } i\in\{1, 2, 3\}\,.
    \label{eq:Ldot} 
\end{equation}

The system (\ref{eq: activeL2} - \ref{eq:Ldot}) can be easily written as a differential-algebraic system of $11$ equations with $11$ unknowns. This system is then solved using standard numerical techniques.


\subsection{Computing net displacement over a period}
\label{sec2:3}

In the four-sphere swimmer described in our model, the net displacement of the swimmer after a periodic stroke can be computed as follows. Notice that the computation below does not involve the nature of the links (whether they are springs or rods).  Calling $X$ the position of the first sphere, $V$ can be expressed linearly in terms of $\dot{X}$ and the rates of change of the arms' length $\dot L_i$ as
$$
v_1 = \dot X,\,v_2 = \dot X + \dot L_1,\,v_3 = \dot X + \dot L_1 + \dot L_2,\,v_4 = \dot X + \dot L_1 + \dot L_2 + \dot L_3\,,
$$
that we rewrite in the matrix form as 
$$
V = M \left(\begin{array}{c}\dot X \\ \dot L_1 \\ \dot L_2 \\ \dot L_3\end{array}\right) \mbox{ with } 
M = \left(\begin{array}{cccc}
1 & 0 & 0 & 0\\1 & 1 & 0 & 0\\1 & 1 & 1 & 0\\1 & 1 & 1 & 1
\end{array}
\right)\,.
$$
Next, using $F=R^{-1} V$ and that the total force vanishes, provides us with a linear relationship between $\dot X$ and $\dot L_1, \dot L_2, \dot L_3$. Namely
\begin{equation}
\begin{pmatrix}
1\\1\\1\\1
\end{pmatrix}\cdot R^{-1}M \begin{pmatrix}\dot X \\ \dot L_1 \\ \dot L_2 \\ \dot L_3\end{pmatrix} =0\,.
\label{eq:motion}
\end{equation}
This is an equation that links $\dot X$ to $\dot L_1, \dot L_2, \dot L_3$ linearly, that we may rewrite under the form
$$
\dot X = \xi(L)\cdot \dot L
$$
where $\xi$ is a 3-D vector field. Indeed, an expansion of $\xi(L)$ around the equilibrium solution $L_{eq}=(l_1,l_2,l_3)$ in the limit $a/l\ll 1$ with $l=\max(l_1,l_2,l_3)$ is obtained as follows. We first notice that, to first order
$$
R^{-1} = 6\pi\mu a \,\text{Id} - 9\pi\mu a\left(
\begin{array}{cccc}
 0 & \ds \frac{a}{L_1} & \ds \frac{a}{L_1+L_2} & \ds \frac{a}{L_1+L_2+L_3} \\
\ds \frac{a}{L_1} & 0 & \ds \frac{a}{L_2} & \ds \frac{a}{L_2+L_3}\\
\ds \frac{a}{L_1+L_2} & \ds \frac{a}{L_2} & 0 & \ds \frac{a}{L_3}\\
\ds \frac{a}{L_1+L_2+L_3} & \ds \frac{a}{L_2+L_3} & \ds \frac{a}{L_3} & 0
\end{array}
\right)\,+ 6\pi\mu a\,O\left(\left(\frac{a}{l}\right)^2\right).
$$
Simplifying by $6\pi\mu a$, we may thus rewrite \eqref{eq:motion} as
$$
\alpha \dot{X} + \beta_1 \dot{L}_1+ \beta_2 \dot{L}_2+ \beta_3 \dot{L}_3 = 0\,,
$$
where 
\begin{eqnarray*}
    \alpha &=& \ds 4 - \frac32\left(\frac{2a}{L_1} + \frac{2a}{L_2} + \frac{2a}{L_3} + \frac{2a}{L_1+L_2}+\frac{2a}{L_2+L_3}+\frac{2a}{L_1+L_2+L_3}\right)+O\left(\left(\frac{a}{l}\right)^2\right)\,,\\
    \beta_1 &=& \ds 3 - \frac32\left(\frac{a}{L_1} + \frac{2a}{L_2} + \frac{2a}{L_3} + \frac{a}{L_1+L_2}+\frac{2a}{L_2+L_3}+\frac{a}{L_1+L_2+L_3}\right)+O\left(\left(\frac{a}{l}\right)^2\right)\,,\\
    \beta_2 &=& \ds 2 - \frac32\left(\frac{a}{L_2} + \frac{2a}{L_3} + \frac{a}{L_1+L_2}+\frac{a}{L_2+L_3}+\frac{a}{L_1+L_2+L_3}\right)+O\left(\left(\frac{a}{l}\right)^2\right)\,,\\
    \beta_3 &=& \ds 1 - \frac32\left(\frac{a}{L_3} + \frac{a}{L_2+L_3}+\frac{a}{L_1+L_2+L_3}\right)+O\left(\left(\frac{a}{l}\right)^2\right)\,,
\end{eqnarray*}
from which we deduce that
$$
\xi = -\frac{(\beta_1,\beta_2,\beta_3)^T}{\alpha} + O\left(\left(\frac{a}{l}\right)^2\right)\,.
$$
This leads to a formula for the coordinates of $\xi$, namely
\begin{eqnarray*}
\xi_1 &=& -\frac34 - \frac{3}{16}\left(\frac{a}{L_1} - \frac{a}{L_2} - \frac{a}{L_3} + \frac{a}{L_1+L_2}-\frac{a}{L_2+L_3}+\frac{a}{L_1+L_2+L_3}\right)+O\left(\left(\frac{a}{l}\right)^2\right)\,,\\
\xi_2 &=& -\frac12 - \frac38\left(\frac{a}{L_1} -\frac{a}{L_3}\right)+O\left(\left(\frac{a}{l}\right)^2\right)\,,\\
\xi_3 &=& -\frac14 - \frac{3}{16}\left(\frac{a}{L_1} + \frac{a}{L_2} - \frac{a}{L_3} + \frac{a}{L_1+L_2}-\frac{a}{L_2+L_3}-\frac{a}{L_1+L_2+L_3}\right)+O\left(\left(\frac{a}{l}\right)^2\right)\,,\\
\end{eqnarray*}

Now, if we consider a closed stroke $L(t)$ in the shape space, the total displacement at the end of the stroke appears to be the circulation of $\xi$ along the closed curve $L(t)$. In other words, using Stokes theorem
$$
\Delta X = \oint_L \xi(L)\cdot dL = \int_\Sigma \mbox{curl}(\xi)\cdot dS
$$
where $\Sigma$ is any surface that is bounded by the closed curve $L(t)$.

From the computation above, the leading term of $\text{curl}(\xi)$ can be computed near $L=(l_1,l_2,l_3)$ and we obtain
\begin{equation}
\text{curl}(\xi)(l_1,l_2,l_3) = \frac{3}{16}\left(\begin{array}{c}
\ds \frac{a}{l_2^2}+\frac{2a}{l_3^2}+\frac{a}{(l_1+l_2)^2}-\frac{a}{(l_2+l_3)^2}-\frac{a}{(l_1+l_2+l_3)^2}\\
\ds -\frac{a}{l_1^2}-\frac{a}{l_3^2}-\frac{a}{(l_1+l_2)^2}-\frac{a}{(l_2+l_3)^2}+\frac{2a}{(l_1+l_2+l_3)^2}\\
\ds \frac{2a}{l_1^2}+\frac{a}{l_2^2}-\frac{a}{(l_1+l_2)^2}+\frac{a}{(l_2+l_3)^2}-\frac{a}{(l_1+l_2+l_3)^2}
\end{array}
\right)+O\left(\frac{a^2}{l^3}\right)\,,
\label{eq:curlxi}
\end{equation}
which enables us to compute an approximation of the displacement of the swimmer after a stroke around a given shape $L=(l_1,l_2,l_3)$. The validity of this expression to compute an approximation of the displacement after a stroke will be checked against the full integration of the system in the next Section. 

\subsection{Numerical results: Swimming dynamics and displacement over time}
\label{sec:2.2}
\begin{figure}[h]
    \centering
    	\begin{subfigure}{0.45\textwidth}
		\centering
		\includegraphics[width=0.9\linewidth]{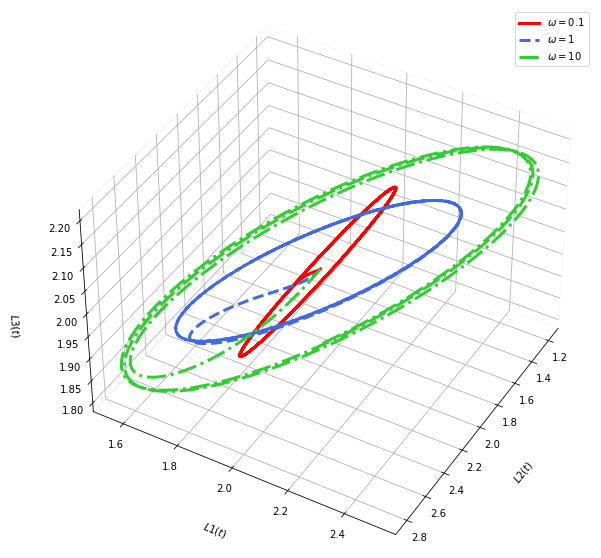}
		\subcaption{}
	\end{subfigure}
	\begin{subfigure}{0.45\textwidth}
		\centering
    \includegraphics[width=0.9\linewidth]{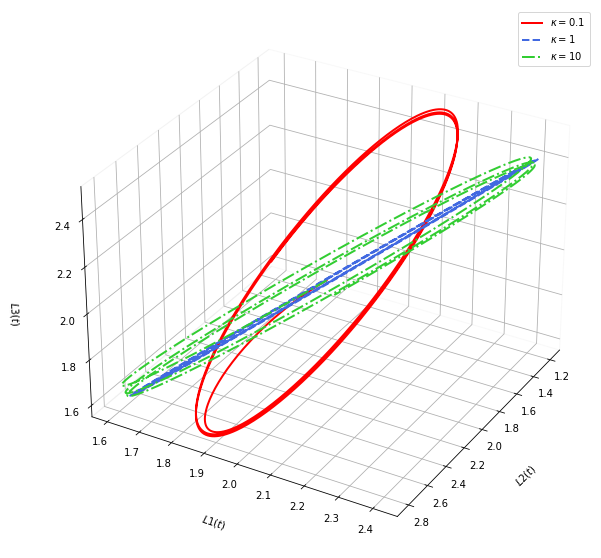}
		\subcaption{}
	\end{subfigure}

    \caption{Three dimensional trajectories of the four-sphere swimmer depicting limit cycles. (A) Effect of various frequencies $\omega$ (in $\unit[]{rad~ s^{-1}}$) for a fixed $\kappa = 10$. (B) Effect of stiffness ratio $\kappa = k_3/k_1$ for a fixed $\omega = \unit[1]{rad~ s^{-1}}$.}
\label{fig:asym_swimmer_traj}
\end{figure}

Numerical integration of the system (\ref{eq: activeL2} - \ref{eq:Ldot}) typically yields the 3D trajectories $(L_{1}(t), L_2(t), L_3(t))$ that are plotted in Figure \ref{fig:asym_swimmer_traj}. The flux of $\text{curl}(\xi)$ through the areas enclosed by the limit cycles represents the displacement of the swimmer per unit cycle. All simulations are performed with reference parameters $l = l_1=l_2= l_3=\unit[2\cdot 10^{-4}]{m}$, $a=\unit[0.1\cdot 10^{-4}]{m}$, $\mu=\unit[8.9\cdot 10^{-4}]{Pa~ s}$, $\varepsilon = 0.4$, $\omega = \unit[1]{rad ~s^{-1}}$. 
As in Section \ref{sec:twolink}, Figure \ref{fig:asym_swimmer_traj}(A) shows the behavior of the system when $k_1 = 0.1 k_3=\unit[10^{-7}]{N ~m^{-1}}$, and for $\omega = \unit[0.1]{rad ~s^{-1}}$, $\unit[1]{rad ~s^{-1}}$, and $\unit[10]{rad ~s^{-1}}$. As expected, a periodic limit cycle is quickly obtained whose orientation changes with $\omega$.  As for the swimmer of the previous section, there exists an optimal frequency that provides the highest displacement per unit cycle, while for very low or high frequencies the area enclosed by the cycle tends to 0.

\begin{figure}[ht!]
\centering
\includegraphics[width=0.6\textwidth]{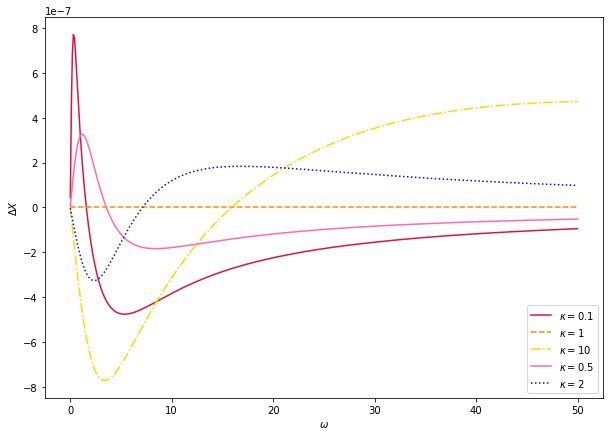}
\caption{Average displacement per period ($\Delta X$, in $m$) of the swimmer, plotted as a function of $\omega$ (in $rad \, s^{-1}$) for various $\kappa$.}
\label{fig:middle_link_displacement_vs_time_global}
\end{figure}

In  Figure \ref{fig:asym_swimmer_traj}(B), we fix the frequency $\omega=1$ and vary the stiffness of the second link relative to the first link. Here $k_3=\unit[1 \cdot 10^{-7}]{N ~m^{-1}}$ while $k_1=0.1 k_3, k_3$ and $10\,k_3$, characterized by the stiffness ratio $ \kappa = k_{3}/k_{1}$.  We observe that, although the areas enclosed by the trajectories remain similar, their orientation varies depending on $\kappa$. This can be expected since, when one spring is stiffer than the other one, phase lag will differ and one side of the swimmer will thus start moving before the other one. This influences the swimming direction of our model, which is then illustrated when computing the displacement.
Lastly, we can notice that the area enclosed by the curve when $\kappa=1$ is nonzero, although the swimmer does not have any net displacement, as shown later on in Figure \ref{fig:middle_link_displacement_vs_time_global}, which seems unusual. This can be nevertheless explained. Indeed, when $\kappa=1$, the swimmer is completely symmetric, falling in the case of the \textit{Scallop Theorem} once again. However, since there is still a phase lag between the springs and the active arm, the global length at which the swimmer starts each stroke can vary, explaining the nonzero surface bounded by the curve in the shape space.

\begin{figure}[ht!]
    \centering
        \includegraphics[width=0.6\textwidth]{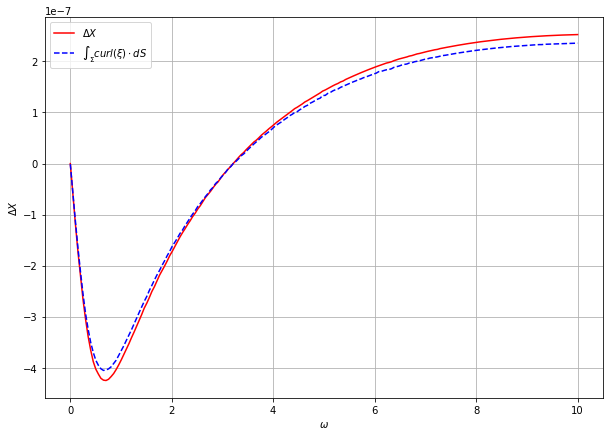}  
    \caption{Average displacement per period ($\Delta X$, in $m$) of the swimmer computed directly by integration (in red) and using the circulation of the first-order approximation \eqref{eq:curlxi} of $\text{curl}(\xi)$ (in blue), plotted as a function of $\omega$ (in $rad \, s^{-1}$), for $\varepsilon = 0.3$.}
\label{fig:deltax_curl}
\end{figure}

We calculate the net displacement over a periodic stroke by integrating the average velocity of the spheres in the steady-state regime, i.e. when the initial transient is terminated. Figure \ref{fig:deltax_curl} shows both the approximation of the displacement computed using the formula $\Delta X = \oint_L \xi(L)\cdot dL = \int_\Sigma \mbox{curl}(\xi)\cdot dS$ with the first-order expansion of $\text{curl}(\xi)$ given in \eqref{eq:curlxi}, and the integration of $\dot X$ giving the exact displacement per period. Here, $a/\ell = 0.05$ and $\varepsilon = 0.3$, and we notice very a very small difference of about $5\%$ between the approximation with the numerical integration of the system over a wide range of frequencies.\\

\begin{figure}[h]
    \centering
        \includegraphics[width=0.6\textwidth]{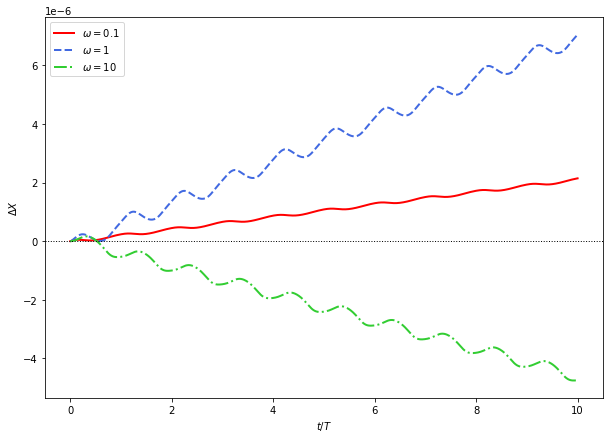}  
    \caption{The swimmer's displacement $\Delta X$ (in $m$) plotted as a function of normalized time $t/T=t\cdot \omega/(2\pi)$ for $\kappa = 0.1$ and various $\omega$.}
\label{fig:middle_link_displacement_vs_time}
\end{figure}

In Figure \ref{fig:middle_link_displacement_vs_time}, we plot the swimmer's displacement as a function of time, in the permanent regime, when $\kappa = k_3/k_1 = 0.1$, for $\omega =  \unit[\{0.1, 1, 10\}]{rad~ s^{-1}}$. The swimmer exhibits different behaviors depending on the frequency, namely, its efficiency varies and, more importantly the direction of the motion changes. 

To understand the variation of the swimmer displacement with the frequency, we plot in Figure \ref{fig:middle_link_displacement_vs_time_global},  the average (renormalized) displacement per period $\Delta X/l$,  as a function of $\omega$. Roughly speaking, the plot segregates two regimes that are separated by the $\kappa=1$ curve. In this latter situation, as we have already mentioned, both springs are synchronized with the same amplitude due to the symmetry of the system. Thus, the swimmer undergoes a reciprocal shape change and no displacement can be observed. For $\kappa>1$, $\Delta X/l$ has a positive maximum first, followed by a negative minimum, and conversely for $\kappa<1$. The average displacement per period approaches to zero as $\omega\to 0$ and $\omega\to\infty$ \cite{montino_three-sphere_2015}. The maxima/minima of the curves imply a characteristic resonant behavior where the displacement is maximum for a given frequency.

Now, consider the case of $\kappa = 0.1$ in Figure \ref{fig:middle_link_displacement_vs_time_global}. The displacement over one stroke $\Delta X$ is maximum at $\omega\approx \unit[0.5]{rad~ s^{-1}}$ and minimum for $\omega\approx \unit[5]{rad~ s^{-1}}$. Therefore, when the right spring is much stiffer than the left spring,  $k_3\gg k_1$, the swimmer moves to the right for small frequencies and moves to the left for moderate frequencies. As $k_3$ reduces and approaches $k_1$, both extrema of $\Delta X$ reduce in magnitude and shift to larger frequencies. For $k_1=k_3$, there is no displacement. For $k_1>k_3$ the scenario is opposite with the minima of larger magnitude appearing first followed by the maxima of smaller amplitude.

\section{Conclusion}
We have proposed and studied the dynamics of a four-sphere swimmer with two passive elastic arms, activated by a central link that elongates and retracts periodically with amplitude $\varepsilon$ and angular frequency $\omega$.
Thanks to interactions between elastic and hydrodynamic forces, a phase difference between the oscillations of the active arm and each spring is created, whose value depends on the springs' stiffness. When these are unequal, it gives rise to a large range of frequencies at which the swimmer moves with a higher efficiency.
This phase differences enables the model to undergo non-reciprocal shape changes, thus circumventing the Scallop Theorem's obstruction \cite{purcell_life_1977}. 

Moreover, since there are two springs with different stiffnesses on each side of the activated arm, depending on the driving frequency, the system undergoes a different behavior. Indeed, playing with the oscillation frequency of the active arm can change the system's swimming direction, contrarily to what was shown in earlier works \cite{montino_three-sphere_2015}. This paves the way for controlling complex trajectories with simple artificial micro-swimmers composed of spheres as in references \cite{Alouges2018, Alouges2020, Alouges2022}.

%
%

\end{document}